\documentclass[conference]{IEEEtran}
\usepackage{times}
\usepackage{graphicx}
\usepackage{cite}
\usepackage{color}
\usepackage{psfrag}
\usepackage{subfigure}
\usepackage{amssymb}
\usepackage{epsfig}
\usepackage{pifont}
\usepackage{amsmath}
\usepackage{array}
\usepackage{multicol}
\usepackage[T1]{fontenc}
\usepackage{comment}
\usepackage{acronym}
\usepackage{multirow}
\usepackage{epsfig}
\begin{document}

\title{HyperRNN: Deep Learning-Aided Downlink CSI Acquisition via Partial Channel Reciprocity for FDD Massive MIMO}
%\author{Yusha Liu and Osvaldo Simeone
\thanks{The authors are with King's Communications, Learning, and Information
Processing (KCLIP) lab at the Department of Engineering of Kings College
London, UK (emails: yusha.liu@kcl.ac.uk, osvaldo.simeone@kcl.ac.uk).
}
\author{\IEEEauthorblockN{Yusha Liu}
\IEEEauthorblockA{\textit{KCLIP Lab, Department of Engineering} \\
\textit{King's College
London},
London, UK \\
yusha.liu@kcl.ac.uk}
\vspace{-0.2 cm}
\and
\IEEEauthorblockN{Osvaldo Simeone}
\IEEEauthorblockA{\textit{KCLIP Lab, Department of Engineering} \\
\textit{King's College
London},
London, UK \\
osvaldo.simeone@kcl.ac.uk}
\vspace{-0.2 cm}}
\maketitle
\begin{abstract}
 In order to unlock the full advantages
of massive multiple input multiple output (MIMO) in the downlink, channel state information (CSI) is required
at the base station (BS) to optimize the beamforming matrices. In frequency division duplex
(FDD) systems, full channel reciprocity does not hold, and CSI acquisition generally requires downlink pilot transmission followed by uplink feedback. Prior work proposed the end-to-end design of pilot transmission, feedback, and CSI estimation via deep learning. In this work, we introduce an enhanced end-to-end design that leverages partial uplink-downlink reciprocity and temporal correlation of the fading processes by utilizing jointly downlink and uplink pilots. The proposed method is based on a novel deep learning architecture -- \emph{HyperRNN} -- that combines hypernetworks and recurrent neural networks (RNNs) to optimize the transfer of long-term channel features from uplink to downlink. Simulation results demonstrate that the HyperRNN achieves a lower normalized mean square error (NMSE) performance, and that it reduces requirements in terms of pilot lengths.
\end{abstract}

\section{Introduction}
%  Obtaining channel state information (CSI) at the transmitter is crucial for the optimization of beaforming matrices in downlink transmission. While in time division duplex (TDD) systems, downlink CSI
% can be computed from uplink CSI thanks to channel reciprocity.
% However, most cellular systems use  protocols,  causing a challenge for downlink CSI acquisition
% at the base station (BS).

With frequency division duplex (FDD), downlink channel state information (CSI) cannot be directly obtained from uplink pilots due to a lack of full reciprocity between uplink and downlink channels. This poses a challenge in massive massive multiple input multiple output (MIMO) systems, since the use of downlink pilots entails a generally large communication overhead to feed back the estimated CSI from users to base station (BS), owing to the massive number of antennas. Solutions to this practically important problem can be divided into uplink training-based, downlink training-based, and hybrid methods. In the first class are schemes that leverage \emph{partial reciprocity} in the form of frequency- and time-invariant multipath parameters, such as angles of arrival/ departure (AoAs/ AoDs) and path gains, to directly map uplink CSI to downlink CSI \cite{arnold2019enabling,alrabeiah2019deep,han2020deep,yang2020deep}. Downlink training-based techniques leverage machine learning for the design of CSI compression and uplink feedback algorithms \cite{wen2018deep,jang2019deep}. Hybrid schemes typically operate sequentially, with the uplink pilots used to identify spatial directions along which to send a reduced number of pilots in the downlink motivated by partial reciprocity \cite{zhang2018directional,khalilsarai2018fdd,han2019fdd}. In this paper, we propose a novel end-to-end design for downlink CSI acquisition that leverages partial uplink-downlink reciprocity and temporal correlation of the fading processes in a hybrid architecture that utilizes the simultaneous transmission of downlink and uplink pilots (see Fig. 1).

The reference downlink-based end-to-end architecture for downlink training, uplink feedback, and channel estimation introduced in \cite{sohrabi2020deep} is illustrated in Fig. 2. In it,  downlink pilots are transmitted by the BS; pilots are processed by a deep neural network (DNN) to produce a feedback message consisting of a given number of bits; and the message is in turn processed by another DNN at the BS for CSI acquisition. Note that reference \cite{sohrabi2020deep} studies also the direct design of downlink beamforming matrices -- a topic that we will cover in an extension of this work. In this paper, we propose an enhanced, hybrid, end-to-end design that  is based on a novel deep learning architecture -- \emph{HyperRNN} -- illustrated in Fig. 3. The main innovation of the approach is that simultaneously transmitted pilot symbols in the uplink, across multiple time slots (see Fig. 1), are leveraged to automatically extract long-term reciprocal channel features via a hypernetwork \cite{ha2016hypernetworks} that determines the weight of the downlink CSI estimation network. Importantly, unlike the existing works reviewed above, such as \cite{zhang2018directional,khalilsarai2018fdd,han2019fdd}, the long-term features are not estimated explicitly, but they implicitly underlie the discriminative mapping implemented by the hypernetwork between uplink pilots and downlink CSI estimation network. The second main innovation is to incorporate recurrent neural networks (RNNs), in lieu of (feedforward) DNNs for both uplink and downlink processing in order to leverage the temporal correlation of the fading amplitudes.

Among other related works, we mention \cite{goutay2020deep}, which proposed the use of hypernetworks for MIMO detection in order to avoid retraining for different channel realizations; as well as  \cite{wang2018deep} that introduced a downlink training-based compression scheme for uplink feedback based on RNNs.

The rest of the paper is organized as follows. Section \ref{sec.System} describes the FDD system model. Section \ref{sec.Downlink CSI estimation} reviews the solution proposed in \cite{sohrabi2020deep}, and Section Section \ref{Hypernetwork} proposes the HyperRNN architecture. Numerical results are provided in Section \ref{sec.Simulation}, and conclusions are provided in  Section \ref{sec.Conclusion}.

\begin{figure*}[ht]
  \centering
    \includegraphics[width=0.84\textwidth]{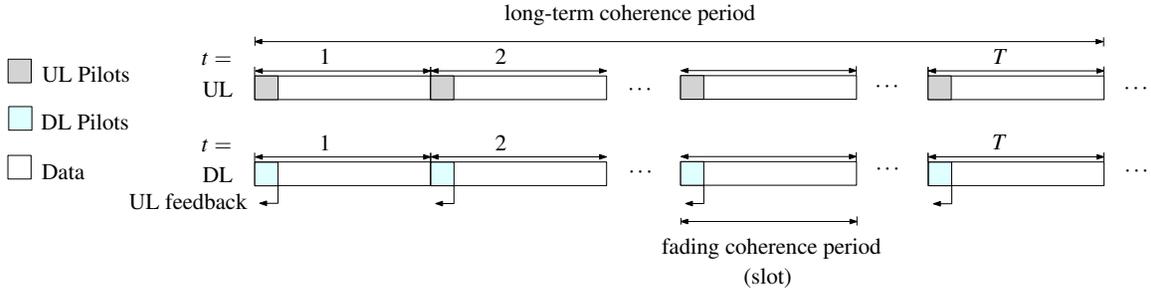}
    \vspace{-0.3 cm}
    \caption{Timeline of uplink and downlink transmission of an FDD massive MIMO system.}
    \vspace{-0.2 cm}
\label{fig.slots}
\end{figure*}

\section{System Model}
\label{sec.System}

We focus on an FDD MIMO system with $M$ transmit antennas (TAs) at BS   supporting a number of single-antenna users. As shown in Fig.~\ref{fig.slots}, since the system uses FDD, uplink and downlink transmissions occur in parallel,  at different carrier frequencies $f_{C}^{\textrm{ul}}$ and  $f_{C}^{\textrm{dl}}$, respectively. We assume orthogonal pilots and no correlation among the channels of different users.
As a result, we can concentrate without loss of generality on a single user.
We study flat-fading channels, which may represent  individual subcarriers in a multi-carrier system.

%\subsection{Uplink Transmission}
%\label{sec.Uplink}
\emph{Uplink Transmission}. The proposed downlink channel estimation framework leverages both uplink and downlink pilots by assuming partial reciprocity.
During uplink transmission, each user transmits $L^{\textrm{ul}}$ uplink pilots $\pmb{x}^{\textrm{ul}} \in \mathbb{C}^{1\times L^{\textrm{ul}}}$ to the BS at the start of the $t$-th  ($t=1,\dots, T$) time slot.
At the BS, the  $M\times L^{\textrm{ul}}$ received discrete-time samples are modelled as
\vspace{-0.1 cm}
\begin{align}\label{eq.yul}
\pmb{Y}_{t}^{\textrm{ul}}=\pmb{h}^{\textrm{ul}}_{t}  \pmb{x}^{\textrm{ul}}  +\pmb{N}_{t}^{\textrm{ul}},
\end{align}
where $\pmb{h}^{\textrm{ul}}_{t} \in \mathbb{C}^{M\times 1}$ is the uplink channel vector and $\pmb{N}_{t}^{\textrm{ul}} \in \mathbb{C}^{M\times L^{\textrm{ul}}}$ is   additive white Gaussian noise (AWGN)  with i.i.d. elements having  zero mean and   variance    $\sigma^2$.
%We assume that the channel is quasi-static, and hence constant within each slot, and that it is characterised by both long-term multipath parameters and fast-varying fading  amplitudes.
After uplink pilot transmission, during the rest of the $t$-th time slot, uplink data symbols  of the user  are transmitted to the BS.

%\subsection{Downlink Transmission}
%\label{sec.Downlink}

%\begin{figure*}[htbp]
%	\centering
%	\subfigure[Uplink transmission]{
%		\centering
%		\includegraphics[width=0.28\linewidth]{figure/uplinktransmission.png}
%		%\label{figBERa}
%	}
%	\subfigure[Downlink transmission]{
%		\centering
%		\includegraphics[width=0.28\linewidth]{figure/downlinktransmission.png}
%		%\label{figBERb}
%	}
%	\caption{FDD massive MIMO system over multipath channels with partial reciprocity: the long-term features $\{\alpha_p,\theta_p\}_{p=1}^{P}$ of the $P$ multipath components are constant and shared between uplink an downlink.}
%	\label{fig.1}
%\end{figure*}
\emph{Downlink Transmission}.
At the same time as the uplink transmission, in the downlink transmission, at each time slot $t=1,\dots, T$, the BS transmits $L^{\textrm{dl}}$ pilot symbols   from the $M$ TAs, which we collect in matrix $\pmb{X}^{\textrm{dl}} \in \mathbb{C}^{M\times L^{\textrm{dl}}}$. The $L^{\textrm{dl}}$ received discrete-time samples at  the user of interest are collected in vector $\pmb{y}_{t}^{\textrm{dl}}\in  \mathbb{C}^{ 1 \times L^{\textrm{dl}}}$, which is modelled as
\vspace{-0.1 cm}
\begin{align}\label{eq.ydl}
\pmb{y}_{t}^{\textrm{dl}}= (\pmb{h}^{\textrm{dl}}_{t})^{H} \pmb{X}^{\textrm{dl}} +\pmb{n}_{t}^{\textrm{dl}},
\end{align}
where $\pmb{h}^{\textrm{dl}}_{t} \in \mathbb{C}^{M\times 1}$ is the downlink channel vector and $\pmb{n}_{t}^{\textrm{dl}}$ is AWGN   with  zero mean and    variance   $\sigma^2$. As for the uplink, we assume a quasi-static channel that is constant within each slot. Upon receiving vector $\pmb{y}_{t}^{\textrm{dl}}$, the user processes it and quantizes the result, producing a message $\pmb{q}^{\textrm{dl}}_{t}=f^{\mathcal{Q}}(\pmb{y}^{\textrm{dl}}_{t})$ of $B$ bits, where $f^{\mathcal{Q}}(\cdot)$ represents the composition of processing and quantization.
The message of $B$ bits is fed back to the BS.

\emph{Channel Model}.
The channel remains constant within a single time slot $t$. Furthermore, we assume standard multipath channels, whereby  each path  is characterized by both long-term features that remain constant for a number $T$ of time slots and fast-varying fading amplitudes \cite{3GPPrelease16}. Specifically, each path $p$, with $p=1,\dots, P$, between user and BS is characterized by the long-term, time-invariant, AoD $\theta_{p}$ and
path gain $\alpha_{p}$; as well as by a time-varying fading process $\beta^{\textrm{ul}}_{p,t}$ for uplink channel and $\beta^{{\textrm{dl}}}_{p,t}$ for downlink channel, respectively.
AoDs $\{\theta_p\}_{p=1}^{P}$ and path gains $\{\alpha_p\}_{p=1}^{P}$ are assumed to be invariant, given the spatial and temporal resolution of the systems, over $T$ time slots. In contrast, the fading processes $\{\beta^{\textrm{ul}}_{p,t}\}_{p=1}^{P}$ and $\{\beta^{\textrm{dl}}_{p,t}\}_{p=1}^{P}$ vary across the time slots $t=1,\dots, T$. Partial reciprocity implies that the long-term multi-path features $\{\theta_p\}_{p=1}^{P}$ and $\{\alpha_p\}_{p=1}^{P}$ are equal for uplink and downlink, while the fading processes are different, being a function of  the carrier frequency $f^{\textrm{ul}}_C$/ $f^{\textrm{dl}}_C$~\cite{han2020deep}.

Given the above assumptions, the $P$-path quasi-static uplink and downlink channel    models of the user  at time slot $t$ are expressed as
\begin{align} \label{channel-ul}
\pmb{h}_{t}^{\textrm{ul}} = \sum_{p=1}^{P}  \alpha_{p}  \pmb{a}^{\textrm{ul}}(\theta_{p})\beta^{\textrm{ul}}_{p,t},~\textrm{and}~\pmb{h}_{t}^{\textrm{dl}}  = \sum_{p=1}^{P}  \alpha_{p}  \pmb{a}^{\textrm{dl}}(\theta_{p})\beta^{\textrm{dl}}_{p,t} ,
\end{align}
where  $\pmb{a}^{\textrm{ul}}( \theta_{p})$ and  $\pmb{a}^{\textrm{dl}}( \theta_{p})$ are the steering vectors, which depend on the antenna array and respective carrier frequencies.
The scaled steering vectors $ \alpha_{p} \pmb{a}^{\textrm{ul}}(\theta_{p})$ and $ \alpha_{p} \pmb{a}^{\textrm{dl}}(\theta_{p})$ are   invariant in time slots $t=1, \dots, T$.
Furthermore, they are related, since they both depend on the AoD $\theta_{p}$, but they are distinct, due to the carrier frequency difference $\Delta f_C = |f_C^{\textrm{ul}} - f_C^{\textrm{dl}}| >0$.

The fading amplitudes $\{\beta^{\textrm{ul}}_{p,t}\}_{t=1}^{T}$ in the uplink and $\{\beta^{\textrm{dl}}_{p,t}\}_{t=1}^{T}$ in the downlink are assumed to be independent and to evolve over the slot index $t$ according to the dynamic model accounting for temporal correlation. As a common example, a first-order
Autoregressive
(AR) model \cite{wang2018deep} can be assumed with correlation
 coefficients   $\rho^{\textrm{ul}} =J_0(2\pi f^{\textrm{ul}}_d \tau) \in [-1, 1]$ and $\rho^{\textrm{dl}} =J_0(2\pi f^{\textrm{dl}}_d \tau)$, where $J_0(\cdot)$ is the zero-order Bessel
function  of the first kind; $\tau$ is the time slot duration; and  $f^{\textrm{dl}}_d = v f^{\textrm{dl}}_C/c$ and $f^{\textrm{ul}}_d = v f^{\textrm{ul}}_C/c$ are the maximum Doppler frequency with $v$ being the mobile velocity.

\section{DNN-based Downlink CSI Estimation}
\label{sec.Downlink CSI estimation}
\begin{figure*}[ht]
  \centering
    \includegraphics[width=0.7\textwidth]{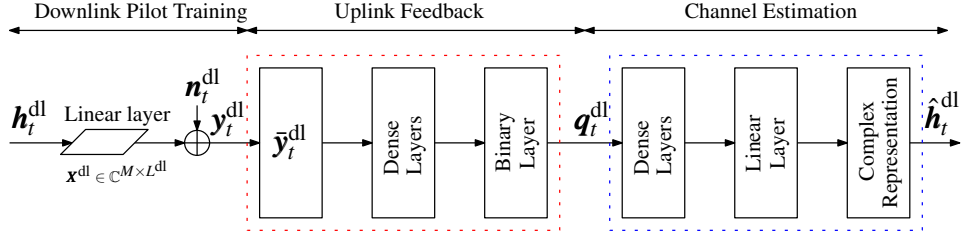}
\vspace{-0.2 cm}    \caption{End-to-end downlink channel estimation based on uplink feedback \cite{sohrabi2020deep}.}
\vspace{-0.1 cm}
\label{fig.2}
\end{figure*}
In this section, we first describe the baseline DNN-based downlink CSI prediction approach  proposed in \cite{sohrabi2020deep}, which estimates the downlink channel $\pmb{h}_{t}^{\textrm{dl}}$ based on the $B$-bit message $\pmb{q}_{t}^{\textrm{dl}}$ obtained from downlink training.
As shown in Fig. \ref{fig.2}, the concatenation of pilot transmission, quantization, and channel estimation is modelled with separate neural network (NN) models and trained in an end-to-end fashion.

\subsubsection{Downlink Pilot Transmission}
To enable end-to-end training, first, the received signal \eqref{eq.ydl}  is modelled as the output of a fully-connected linear layer with input $\pmb{h}^{\textrm{dl}}_{t}$, weight matrix $\pmb{X}^{\textrm{dl}}$, and output $\pmb{y}_{t}^{\textrm{dl}}=\pmb{h}^{\textrm{dl}}_{t}\pmb{X}^{\textrm{dl}}$, to which Gaussian noise $\pmb{n}_{t}^{\textrm{dl}}$ is added.  The training matrix $\pmb{X}^{\textrm{dl}}$ is subject to design.

\subsubsection{Uplink  Feedback Quantization}
In order to produce the $B$-bit message $\pmb{q}^{\textrm{dl}}_{t}$, upon receiving the signal $\pmb{y}_{t}^{\textrm{dl}}$ in \eqref{eq.ydl},  the user applies a multi-layer fully-connected DNN with a sign$(\cdot)$ activation function at the last layer. Specifically, as shown in Fig.~\ref{fig.2}, the inputs  of the DNN are comprised of the real and imaginary parts of elements in $\pmb{y}^{\textrm{dl}}_{t}$, which can be expressed as the $2L^{\textrm{dl}} \times 1$ vector
%\begin{align}
$\bar{\pmb{y}}^{\textrm{dl}}_{t} = \textrm{c2r} (\pmb{y}^{\textrm{dl}}_{t} ) = \left[\Re(\pmb{y}_{t}^{\textrm{dl}})^T, \Im(\pmb{y}_{t}^{\textrm{dl}})^T \right]^T$,
%\end{align}
where the function $\textrm{c2r} (x)= [\Re(x)^T~ \Im(x)^T]^T$ denotes the complex  to real value representation, with $\Re(\cdot)$ and $\Im(\cdot)$ representing the real and imaginary parts of the elements in $\pmb{y}^{\textrm{dl}}_{t}$, respectively. Then, vector $\bar{\pmb{y}}^{\textrm{dl}}_{t}$ is processed through $(M^{\mathcal{Q}}-1)$ fully connected layers  with rectified linear unit (ReLU) activation functions, while the last layer produces $B$ binary outputs through the mentioned sign$(\cdot)$ non-linearity.

Accordingly, denoting $\ell_m^{\mathcal{Q}}$  as the number of ReLU neurons in the $m$-th layer, the optimization parameters of the quantization DNN are
${\pmb{\Omega}}^{\mathcal{Q}} = \{{\pmb{W}}_{1}^{\mathcal{Q}}, {\pmb{b}}_{1}^{\mathcal{Q}}, \cdots, {\pmb{W}}_{M^{\mathcal{Q}}}^{\mathcal{Q}}, {\pmb{b}}_{M^{\mathcal{Q}}}^{\mathcal{Q}}\}$, with $\ell_m^{\mathcal{Q}}\times \ell_{m+1}^{\mathcal{Q}}$ weight matrix ${\pmb{W}}_{m}^{\mathcal{Q}}$ and $\ell_m^{\mathcal{Q}}\times 1$  bias vector ${\pmb{b}}_{m}^{\mathcal{Q}}$, with $\ell_{1}^{\mathcal{Q}}=2L^{\textrm{dl}}$ and $\ell_{M^{\mathcal{Q}}+1}^{\mathcal{Q}}=B$.
%The binary outputs of the DNN are given as
%\begin{align}
%\pmb{q}_{t}^{\textrm{dl}}=&\textrm{sign}\left({\pmb{W}}_{M^{\mathcal{Q}}}^{\mathcal{Q}}\left[\cdots f_{\text{ReLU}}\left(\pmb{W}_{1}^{\mathcal{Q}}\bar{\pmb{y}}_t^{\textrm{dl}} +\pmb{b}_{1}^{\mathcal{Q}}\right) \cdots \right]+\pmb{b}_{M^{\mathcal{Q}}}^{\mathcal{Q}}\right) \nonumber \\
% \triangleq & f^{\mathcal{Q}} \left (\pmb{y}^{\textrm{dl}}_t| \pmb{\Omega}^{\mathcal{Q}} \right ),
%\end{align}
%where the $\textrm{sign}(\cdot)$  activation function of the binary layer and the  ReLU  function $f_{\text{ReLU}}(\cdot)=\max(x,0)$ are applied elementwise.

\subsubsection{Channel Estimation}
The output $\pmb{q}_{t}^{\textrm{dl}}$ of the quantization DNN at the user is forwarded to the DNN that is employed for downlink CSI estimation at the BS.
 The  $B \times 1$ vector $\pmb{q}_{t}^{\textrm{dl}}$ is processed through $M^{\mathcal{E}}$ fully connected layers. The first $(M^{\mathcal{E}}-1)$ ReLU hidden layers have $\ell_m^{\mathcal{E}}$ neurons, with $m=1, \dots, M^{\mathcal{E}}-1$; while the last linear layer produces $2M$ real-valued outputs.
The optimization parameters ${\pmb{\Omega}}^{\mathcal{E}}$  for  the  channel estimation  DNN  are
${\pmb{\Omega}}^{\mathcal{E}} = \{{\pmb{W}}_{1}^{\mathcal{E}}, {\pmb{b}}_{1}^{\mathcal{E}}, \dots, {\pmb{W}}_{M^{\mathcal{E}}}^{\mathcal{E}}, {\pmb{b}}_{M^{\mathcal{E}}}^{\mathcal{E}}\}$, where ${\pmb{W}}_{m}^{\mathcal{E}}$ is the $\ell_m^{\mathcal{E}}\times \ell_{m+1}^{\mathcal{E}}$ weight matrix  and ${\pmb{b}}_{m}^{\mathcal{E}}$ represents $\ell_m^{\mathcal{E}}\times 1$  bias vector, with $\ell_{1}^{\mathcal{E}}=B$ and $\ell_{M^{\mathcal{E}}+1}^{\mathcal{E}}=2M$.
%The output of the channel estimation DNN is accordingly expressed as
%\begin{align}
%  \hat{\pmb{h}}_{t}^{\textrm{dl}}&=  \pmb{W}_{M^{\mathcal{E}}}^{\mathcal{E}}
% \left[\cdots    f_{\text{ReLU}}\left(\pmb{W}_{1}^{\mathcal{E}}
%\pmb{q}_{t}^{\textrm{dl}} +\pmb{b}_{1}^{\mathcal{E}}\right) \cdots \right]
% +\pmb{b}_{N}^{\mathcal{E}},\nonumber \\
%&  \triangleq f^{\mathcal{E}}\left (\pmb{q}^{\textrm{dl}}_{t}| \pmb{\Omega}^{\mathcal{E}} \right ).
%\end{align}

\subsubsection{Training}
End-to-end training of the pilot matrix $\pmb{X}^{\textrm{dl}}$ and of the DNN parameters $\pmb{\Omega}^{\mathcal{Q}}$ and $\pmb{\Omega}^{\mathcal{E}}$ is done by minimizing the
training  squared error   between  the DNN output $\hat{\pmb{h}}_{t}^{\textrm{dl}}$ and the real ${\pmb{h}}_{t}^{\textrm{dl}}$
%\begin{subequations}\label{eq.pf}
%\begin{align} \label{pf-1}
%    \underset{\pmb{X}^{\textrm{dl}}, \pmb{\Omega}^{\mathcal{Q}},\pmb{\Omega}^{\mathcal{E}} }{\text{min}}&  \textrm{E} \left [ \|\hat{\pmb{h}}_{t}^{\textrm{dl}} - {\pmb{h}}_{t}^{\textrm{dl}}\|^2 \right ]  \\
%    \text{s.t.}~ &\hat{\pmb{h}}_{t}^{\textrm{dl}}= f^{\mathcal{E}}\left (\pmb{q}^{\textrm{dl}}_{t}| \pmb{\Omega}^{\mathcal{E}} \right ) ,  \\
%    &\pmb{q}^{\textrm{dl}}_{t}=f^{\mathcal{Q}}\left (\pmb{y}_t^{\textrm{dl}}| \pmb{\Omega}^{\mathcal{Q}} \right ),  \\
%    %&\text{Tr}(\pmb{V}\pmb{V}^H)\leq P,\\
%    \label{eq.Pdlconstraint}
%    &\|\pmb{X}^{\textrm{dl}}_{l}\|^2\leq P^{\textrm{dl}}, \forall l = 1,\cdots, L^{\textrm{dl}},
%\end{align}
%\end{subequations}
%where $\pmb{X}^{\textrm{dl}}_{l}$ represents the $l$-th column of $\pmb{X}^{\textrm{dl}}$, and $P^{\textrm{dl}}$ is the transmit power constraint at   BS.
%Problem \eqref{eq.pf} is approximated
by using an empirical average over the training sample set $\{\pmb{y}_{t}^{\textrm{dl}}, \pmb{h}_{t}^{\textrm{dl}}\}$ in lieu of the expectation. %in  \eqref{pf-1}.
%In order to satisfy the transmit power constraint of \eqref{eq.Pdlconstraint}, we employ a weight constraint for each column $l$ of $\pmb{X}^{\textrm{dl}}$, ensuring that $\|\pmb{X}^{\textrm{dl}}_{l}\|^2\leq P^{\textrm{dl}}$.
%The implementation details will be discussed in Section~\ref{sec.Implementation1}. %When implementing the downlink pilot optimization in  TensorFlow

\section{HyperRNN-Based Downlink CSI Estimation}
\label{Hypernetwork}

\begin{figure*}[ht]
  \centering
    \includegraphics[width=0.83\textwidth]{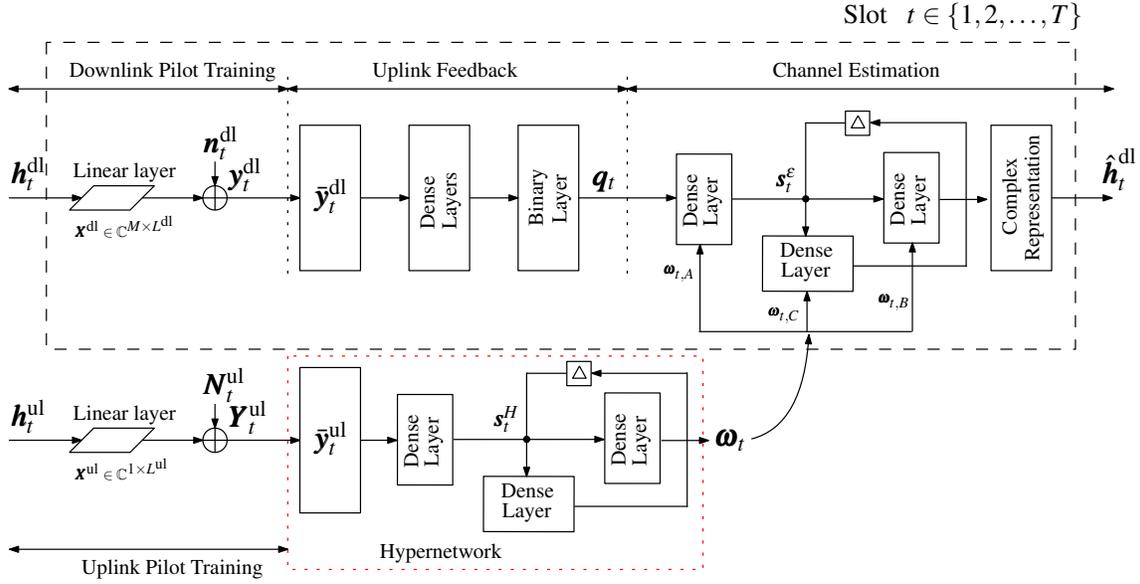}
    \vspace{-0.2 cm}
    \caption{The proposed HyperRNN architecture for end-to-end channel estimation based on temporal correlations and partial reciprocity.}
    \vspace{-0.3 cm}
\label{fig.5}
\end{figure*}
%The approach proposed in \cite{sohrabi2020deep} and reviewed above has two main limitations that we aim at addressing. First, the method estimates the downlink channel  across different time slots $t$ independently, hence not capturing the long-term invariance of the angles and path gains and the temporal correlation of the fading process. Second, the partial reciprocity of uplink and downlink channels is not taken into account. By contrast, the proposed HyperRNN scheme builds on the idea of modelling both temporal correlation and partial reciprocity.

As shown in Fig. \ref{fig.5}, the proposed HyperRNN architecture aims at extracting invariant  reciprocal   features from multipath uplink pilot transmissions by integrating a hypernetwork that takes the input of the signals received in the uplink and outputs the weights of the channel estimation RNN \cite{ha2016hypernetworks}. The key idea is that the weights of the channel estimation RNN can encode information about the long-term channel features that are common to both uplink and downlink, hence automatically accounting for partial reciprocity.

\subsubsection{Uplink Pilot Transmission}

As in the downlink, uplink pilot transmission process is modelled by a a fully-connected linear layer with  input being $\pmb{h}^{\textrm{ul}}_{t}$, weight matrix being $\pmb{x}^{\textrm{ul}}$, and output being $\pmb{Y}_{t}^{\textrm{ul}}$, to which Gaussian noise $\pmb{N}_{t}^{\textrm{ul}}$ is added.

\subsubsection{Downlink Pilot Transmission, Uplink Feedback, and Channel Estimation}
Downlink pilot transmission and uplink feedback quantization operate as discussed in Section \ref{sec.Downlink CSI estimation}, with the caveat that
 in order to leverage the long-term invariance and short-term time-correlation properties of the downlink channel, we replace the channel estimation DNN with an RNN. At each time slot $t$, the RNN produces the estimate
\begin{align} \label{eq.RNN-2}
\textrm{c2r} (\hat{\pmb{h}}_{t}^{\textrm{dl}})= \pmb{W}^{\mathcal{E}}_B  \pmb{s}^{\mathcal{E}}_{t} + \pmb{b}^{\mathcal{E}}_{B}
\triangleq  f^{\mathcal{E}} \left(\pmb{q}_{t}^{\textrm{dl}},\pmb{s}^{\mathcal{E}}_{t-1} | \pmb{\Omega}^{\mathcal{E}}  \right),
\end{align}
where the $\ell^{\mathcal{E}} \times 1$ internal state of the RNN evolves as
\vspace{-0.1 cm}
\begin{align} \label{eq.RNN-3}
 \pmb{s}^{\mathcal{E}}_{t}
 &= f_{\textrm{ReLU}}\left( \pmb{W}^{\mathcal{E}}_A   \pmb{q}^{\textrm{dl}}_t+\pmb{W}^{\mathcal{E}}_C  \pmb{s}^{\mathcal{E}}_{t-1}+ \pmb{b}^{\mathcal{E}}_{A} \right),
\end{align}
with $\pmb{W}^{\mathcal{E}}_A  \in \mathbb{C}^{\ell^{\mathcal{E}}\times B} $,  $\pmb{b}^{\mathcal{E}}_{A}\in \mathbb{C}^{\ell^{\mathcal{E}}\times 1}$,
$\pmb{W}^{\mathcal{E}}_B  \in \mathbb{C}^{2M\times\ell^{\mathcal{E}}}$,  $\pmb{b}_{B}\in \mathbb{C}^{2M\times1}$, $\pmb{W}^{\mathcal{E}}_C  \in \mathbb{C}^{\ell^{\mathcal{E}}\times \ell^{\mathcal{E}}}$,
$\pmb{s}^{\mathcal{E}}_{t}\in \mathbb{C}^{\ell^{\mathcal{E}}\times 1}$ and $\ell^{\mathcal{E}}$ denoting the number of neurons employed for the fully-connected ReLU layer.

\subsubsection{HyperRNN}

To leverage partial reciprocity, we introduce a hypernetwork \cite{ha2016hypernetworks} to adjust the weights of the  downlink channel estimation RNN based on the uplink received signal $\pmb{Y}_{t}^{\textrm{ul}}$, as shown in Fig. \ref{fig.5}.
In order to reduce the number of outputs of the hypernetwork, as in, e.g., \cite{goutay2020deep}, the hypernetwork   generates a common scaling factor for each column of the weight matrices at each time slot $t$. Accordingly,  the output of the hypernetwork is a $(B+2\ell^{\mathcal{E}})\times 1$ vector $\pmb{\omega}_t$, which is detailed next.

The real and imaginary parts of the received uplink signal $\pmb{Y}_{t}^{\textrm{ul}}$ are collected in the $2ML^{\textrm{ul}}\times 1$ vector $\bar{\pmb{y}}_{t}^{\textrm{ul}}=  \textrm{c2r}(\textrm{vec}(\pmb{Y}_{t}^{\textrm{ul}}) )$, with $\textrm{vec}(\cdot)$ denoting the vectorization of a matrix by stacking columns. This vector is fed as
 input to the hypernetwork, together with the internal state $\pmb{s}^{\mathcal{H}}_{t-1}$  from the previous time slot $t-1$. In a manner similar to \eqref{eq.RNN-2}-\eqref{eq.RNN-3}, the
hypernetwork operates as
\vspace{-0.1 cm}
\begin{subequations}\label{eq.RNN}
\begin{align} \label{RNN-1}
\pmb{s}^{\mathcal{H}}_{t}=& f_{\textrm{ReLU}}\left( \pmb{W}^{\mathcal{H}}_A \bar{\pmb{y}}^{\textrm{ul}}_t+\pmb{W}^{\mathcal{H}}_C\pmb{s}^{\mathcal{H}}_{t-1}+ \pmb{b}^{\mathcal{H}}_{A} \right)\\
\textrm{and} ~~~\pmb{\omega}_t=&\pmb{W}^{\mathcal{H}}_B \pmb{s}^{\mathcal{H}}_{t} + \pmb{b}^{\mathcal{H}}_{B}
\triangleq f^{\mathcal{H}} \left(\pmb{y}^{\textrm{ul}}_t,\pmb{s}^{\mathcal{H}}_{t-1} |  \pmb{\Omega}^{\mathcal{H}}  \right),
\end{align}
\end{subequations}
where $\pmb{s}^{\mathcal{H}}_{t}\in \mathbb{C}^{\ell^{\mathcal{H}}\times 1}$ is the internal state, and  $\pmb{W}^{\mathcal{H}}_A  \in \mathbb{C}^{\ell^{\mathcal{H}}\times 2ML^{\textrm{ul}}} $,  $\pmb{b}^{\mathcal{H}}_{A}\in \mathbb{C}^{\ell^{\mathcal{H}}\times 1}$,
$\pmb{W}^{\mathcal{H}}_B  \in \mathbb{C}^{(B+ 2\ell^{\mathcal{E}})\times\ell^{\mathcal{H}}} $,
$\pmb{b}^{\mathcal{H}}_{B}\in \mathbb{C}^{(B+ 2\ell^{\mathcal{E}})\times1}$,
$\pmb{W}^{\mathcal{H}}_C  \in \mathbb{C}^{\ell^{\mathcal{H}}\times \ell^{\mathcal{H}}}$ are optimization parameters  $\pmb{\Omega}^{\mathcal{H}} = \{{\pmb{W}}_{A}^{\mathcal{H}}, {\pmb{b}}_{A}^{\mathcal{H}}, {\pmb{W}}_{B}^{\mathcal{H}}, {\pmb{b}}_{B}^{\mathcal{H}}, {\pmb{W}}_{C}^{\mathcal{H}}\}$ of the hypernetwork.
The $(B+ 2\ell^{\mathcal{E}}) \times 1$ output vector $\pmb{\omega}_t$ modifies the weights of  the downlink channel estimation  RNN \eqref{eq.RNN-2}-\eqref{eq.RNN-3} as
%\begin{subequations}\label{eq.weight}
%\begin{align} \label{weight-1}
$\pmb{W}^{\mathcal{E}}_A = \bar{\pmb{W}}^{\mathcal{E}}_A \cdot \textrm{diag} \left \{ \pmb{\omega}_{t,A} \right \}$,
$\pmb{W}^{\mathcal{E}}_B = \bar{\pmb{W}}^{\mathcal{E}}_B \cdot \textrm{diag} \left \{ \pmb{\omega}_{t,B} \right \}$, and
$\pmb{W}^{\mathcal{E}}_C = \bar{\pmb{W}}^{\mathcal{E}}_C \cdot \textrm{diag} \left \{ \pmb{\omega}_{t,C} \right \}$.
%\end{align}
%\end{subequations}
We have partitioned the output of the hypernetwork as $\pmb{\omega}_t= [\pmb{\omega}_{t,A}, \pmb{\omega}_{t,B},\pmb{\omega}_{t,C}]$, where $\pmb{\omega}_{t,A} \in \mathbb{C}^{B\times 1}, \pmb{\omega}_{t,B} \in \mathbb{C}^{\ell^{\mathcal{E}}\times 1}$, and $\pmb{\omega}_{t,C} \in \mathbb{C}^{\ell^{\mathcal{E}}\times 1}$.
The matrices $\bar{\pmb{W}}^{\mathcal{E}}_A \in \mathbb{C}^{\ell^{\mathcal{E}}\times B}$, $\bar{\pmb{W}}^{\mathcal{E}}_B \in \mathbb{C}^{2M\times\ell^{\mathcal{E}}}$ and $\bar{\pmb{W}}^{\mathcal{E}}_C \in \mathbb{C}^{\ell^{\mathcal{E}}\times \ell^{\mathcal{E}}}$ are also subject to optimization, but, unlike vector $\pmb{\omega}_t$, they are fixed at run time and they are not adapted to the received signals. Therefore, the matrices $\bar{\pmb{W}}^{\mathcal{E}}_A$, $\bar{\pmb{W}}^{\mathcal{E}}_B$ and $\bar{\pmb{W}}^{\mathcal{E}}_C$ cannot account for the specific long-term features of the channel in the current frame of $T$ time slots. We define ${\pmb{\Omega}}^{\mathcal{E}}= \{\bar{\pmb{W}}_{A}^{\mathcal{E}}, {\pmb{b}}_{A}^{\mathcal{E}}, \bar{\pmb{W}}_{B}^{\mathcal{E}}, {\pmb{b}}_{B}^{\mathcal{E}}, \bar{\pmb{W}}_{C}^{\mathcal{E}}\}$ as the set of optimization parameters for the channel estimation HyperRNN.

\subsubsection{Training}
The  proposed HyperRNN architecture is also trained using an end-to-end approach that aims at minimizing the training squared error between the real and estimated channel. The corresponding optimization problem can be formulated as
\vspace{-0.1 cm}
\begin{subequations}\label{eq.pfRNN}
\begin{align} \label{eq.pfRNN-1}
    \underset{\pmb{x}^{\textrm{ul}},\pmb{X}^{\textrm{dl}}, \pmb{\Omega}^{\mathcal{Q}},\pmb{\Omega}^{\mathcal{H}}, \pmb{\Omega}^{\mathcal{E}} }{\text{min}}& \sum_{t=1}^{T} E \left [ \|\hat{\pmb{h}}_{t}^{\textrm{dl}} - {\pmb{h}}_{t}^{\textrm{dl}}\|^2 \right ]  \\
    \text{s.t.}~
    &\|\pmb{X}^{\textrm{dl}}_{l}\|^2\leq P^{\textrm{dl}},~\forall l = 1,\dots, L^{\textrm{dl}},\\
    &|x_l^{\textrm{ul}}|^2\leq P^{\textrm{ul}}, ~\forall l = 1,\dots, L^{\textrm{ul}},
\end{align}
\end{subequations}
where  $P^{\textrm{dl}}$ and $P^{\textrm{ul}}$ are the transmit power constraint at   the BS and at the user side, respectively. $\pmb{X}^{\textrm{dl}}_{l}$ represents the $l$-th column of $\pmb{X}^{\textrm{dl}}$, while $x_l^{\textrm{ul}}$ is the $l$-th element in $\pmb{x}^{\textrm{ul}}$.
%In contrast to \eqref{eq.pf}, problem \eqref{eq.pfRNN} encompasses the optimizations over the parameter $\pmb{\Omega}^{\mathcal{H}}$  of the hypernetwork, as well as over the uplink pilots $\pmb{X}^{\textrm{ul}}$. Furthermore, the objective in \eqref{eq.pfRNN-1} accounts for the hypernetwork on a frame of $T$ time slots corresponding to a long-term coherence period.
The empirical distribution of a training sample set  $\{\pmb{y}_{t'}^{\textrm{dl}}, \pmb{h}_{t'}^{\textrm{dl}}, \pmb{Y}_{t'}^{\textrm{ul}}, \pmb{h}_{t'}^{\textrm{ul}}\}_{t'=1}^{T}$ is used to  approximate  the expectation in \eqref{eq.pfRNN-1}.

%\vspace{-0.1 cm}
\section{System Performance}
\label{sec.Simulation}

In this section, we  characterise the performance of the proposed HyperRNN for channel estimation.

% and sum rate for beamforming design in Section \ref{sec.Implementation2}.
%\subsection{Implementation Details}
%\label{sec.Implementation1}
\emph{Implementation Details.} In this paper, we employ the  spatial channel model (SCM) standardized in 3GPP Release 16 \cite{3GPPrelease16}, with the simulation parameters summarised in Table \ref{tab.parameter}.
The proposed HyperRNN is implemented using the standard deep learning libraries  TensorFlow and Keras, and we adopt the  adaptive moment estimation (Adam) optimizer with the mini-batch size of 1024 and  a learning rate gradually decreasing from $10^{-3}$ to $10^{-5}$.
For the uplink feedback DNN, $M^{\mathcal{Q}}=4$ dense layers are employed, with $\ell^{\mathcal{Q}}_1=1024$, $\ell^{\mathcal{Q}}_2=512$, $\ell^{\mathcal{Q}}_3=256$, and $\ell^{\mathcal{Q}}_4=B$ ReLU hidden neurons. Furthermore, the RNN   for channel estimation  employs $\ell^{\mathcal{E}}= \ell^{\mathcal{P}}= 256$ ReLU neurons for each hidden layer, whereas $\ell^{\mathcal{H}}= 1024$ ReLU hidden neurons are used for the hypernetwork.
In order to satisfy the power constraint, we normalise the updated $\pmb{x}^{\textrm{ul}}$ or $\pmb{X}^{\textrm{dl}}$ in each iteration to ensure $| x_l^{\textrm{ul}}|^2= P^{\textrm{ul}}$ or $\| \pmb{X}^{\textrm{dl}}_l\|^2= P^{\textrm{dl}}$.
We use the normalized mean square error (NMSE) to characterise the channel estimation performance, which is calculated as
$\textrm{NMSE}  = E   [ \|\hat{\pmb{h}}_{t}^{\textrm{dl}} - {\pmb{h}}_{t}^{\textrm{dl}}\|^2/ \|{\pmb{h}}_{t}^{\textrm{dl}}\|^2].$

\begin{table}[tbp]
\centering
%\arraystretch{1.05}
%\color{blue}
\caption{Simulation parameters}
\vspace{-0.1 cm}
\begin{tabular}{l||r}
\hline
Parameters   &  Values
 \\ \hline
Uplink carrier frequency ($f_C^{\textrm{ul}}$) &   3  GHz \\ \hline
Carrier frequency difference ($\Delta f_C$) &   100 MHz \\ \hline
Mobile velocity ($v$) & 30 km/h \\ \hline
Time slot duration ($\tau$) & $0.1$ ms\\ \hline
No. of paths ($P$) &   $2, 4, 8, 16$ \\ \hline
No. of TAs at the BS ($M$) &  64 \\ \hline
AoDs  ($\theta _p$) &  $\theta _p\sim \mathcal{U}(-\pi/6,\pi/6)$\\ \hline
No. of uplink feedback bits ($B$)&  $[5,30]$ \\ \hline
No. of downlink pilots ($L^{\textrm{dl}}$) &  $2$ \\ \hline
No. of uplink pilots ($L^{\textrm{ul}}$) &  $1, 2, 4$ \\ \hline
%Learning rate &  $10^{-3}$ to $10^{-5}$ \\ \hline
%Mini-batch size &  $1024$ \\ \hline
Signal-to-noise ratio (SNR) & 10 dB \\ \hline
\end{tabular}
\vspace{-0.21 cm}
\label{tab.parameter}
\end{table}

\begin{figure}[t]
  \centering
    \includegraphics[width=0.49\textwidth]{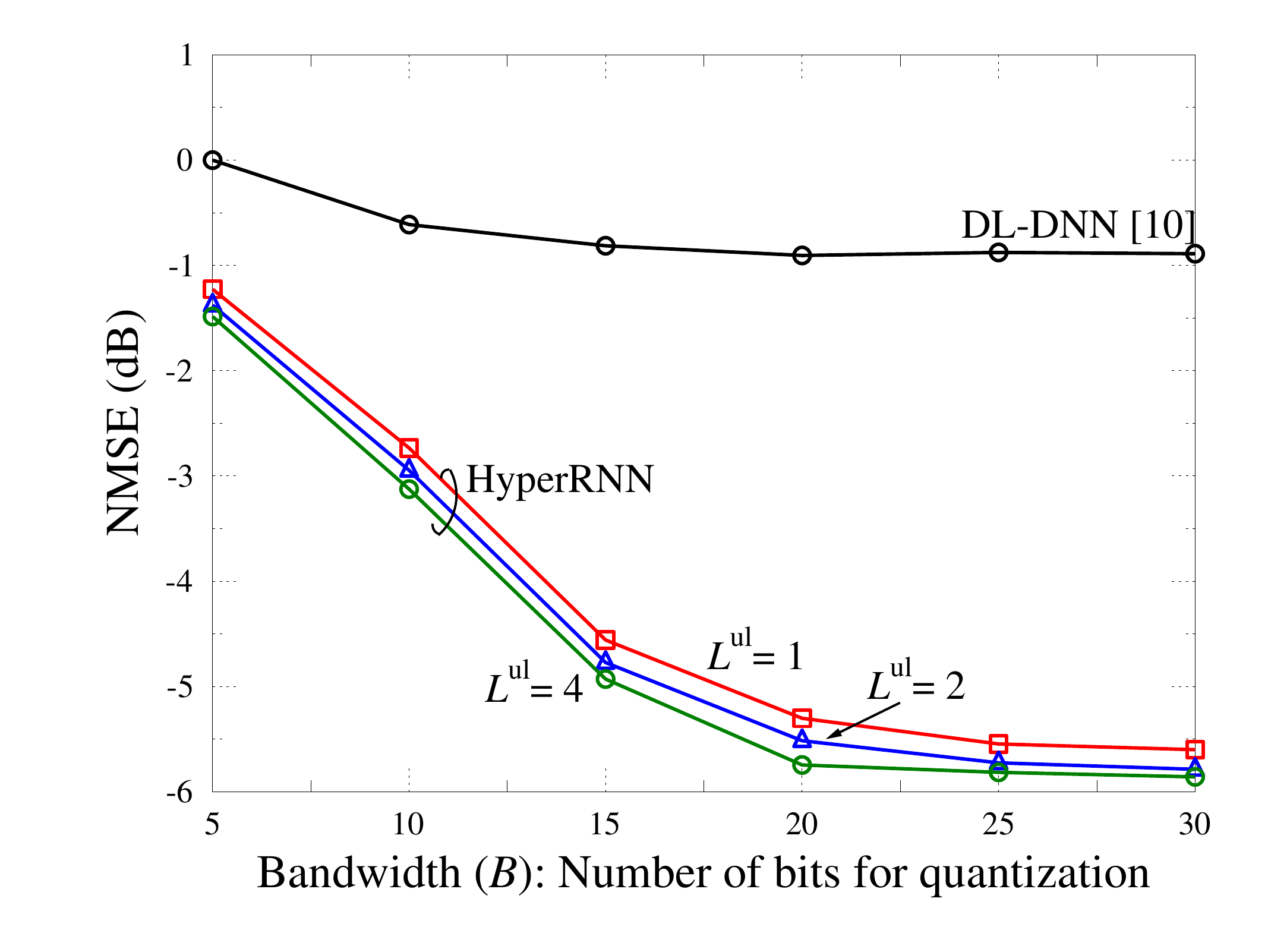}
\vspace{-0.6 cm}
\caption{NMSE of the HyperRNN and DL-DNN \cite{sohrabi2020deep} for an FDD  system with $M=64$  at $t=8$-th time slot, with different uplink pilot lengths $L^{\textrm{ul}}=1, 2$ and 4  over frequency-flat   fading channels, where $L^{\textrm{dl}}=2$, $P=8$, $\Delta f_C = 100$ MHz, and $\rho^{\textrm{ul}}= \rho^{\textrm{dl}}=0$.}
    \vspace{-0.4 cm}
\label{fig.6}
\end{figure}

We compare the NMSE of the proposed HyperRNN for downlink channel estimation with the baseline method DL-DNN proposed in \cite{sohrabi2020deep} using different uplink pilot lengths. In order to isolate the advantage of leveraging long-term partial reciprocity extracted from the uplink, we assume i.i.d. fading amplitudes $\beta_{p,t}^{\textrm{ul}}$ and $\beta_{p,t}^{\textrm{ul}}$ over different time slots in this experiment, i.e., we set the temporal correlation coefficients as $\rho^{\textrm{ul}}= \rho^{\textrm{dl}}=0$. We evaluate the NMSE at the $t=8$-th time slot.
Fig. \ref{fig.6} show that long-term partial reciprocity can be leveraged to enhance   channel estimation, even if a very short uplink pilot sequence length with $L^{\textrm{ul}}=1$ is considered. When a longer pilot sequence is employed, for example, $L^{\textrm{ul}}=4$ the NMSE performance of HyperRNN is improved.
The NMSE reduction is particularly pronounced for longer values of the uplink feedback resolution $B$. This is expected since a longer value of $B$
increases the input size to the downlink channel estimation RNN, increasing the dimension of the output of the hypernetwork.
\begin{figure}[t]
  \centering
    \includegraphics[width=0.49\textwidth]{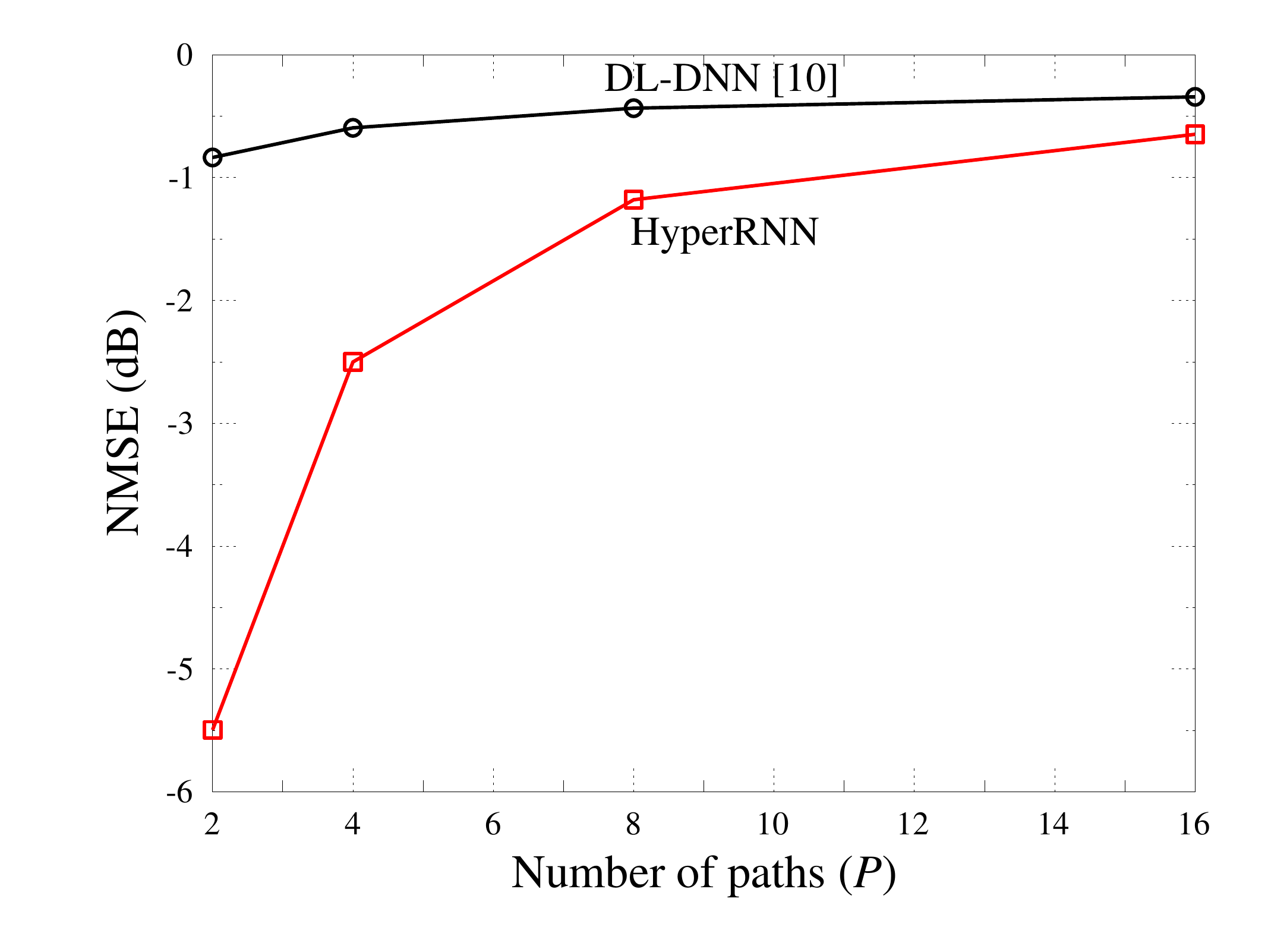}
    \vspace{-0.6 cm}
    \caption{NMSE of the HyperRNN and DL-DNN \cite{sohrabi2020deep} over frequency-flat   fading channels having different number $P$ of paths for an $M=64$ FDD system with $B=20$, ${L}^{\textrm{ul}}={L}^{\textrm{dl}}=2$  and $\Delta f_C = 100$ MHz,  and $\rho^{\textrm{dl}}=0.99$.}
    \vspace{-0.45 cm}
\label{fig.7}
\end{figure}

Fig. \ref{fig.7} demonstrates the NMSE of the proposed  HyperRNN and of the benchmark DL-DNN  \cite{sohrabi2020deep} for channel estimation of the FDD system having different number of paths, $P$, where  $t=8$, ${L}^{\textrm{ul}}={L}^{\textrm{dl}}=2$,  and $\Delta f_C = 100$ MHz,   and the rest parameters are summarised in Table \ref{tab.parameter}.  Note that the temporal correlations $\rho^{\textrm{ul}}$  and $\rho^{\textrm{dl}}$ are not zero here. Larger performance gains can be achieved when  the channel has a lower number of paths. In fact, in this regime, the invariant of the long-term features of the channel defines a low-rank structure of the channel that can be leveraged by the hypernetwork.

\vspace{-0.05 cm}
\section{Conclusions}
\label{sec.Conclusion}

In this paper, we have introduced an end-to-end trained CSI acquisition scheme for massive MIMO FDD systems based on a novel HyperRNN architecture that leverages both partial uplink-downlink reciprocity and temporal correlation of fading processes. The proposed HyperRNN achieves a lower NMSE compared to existing methods, particularly in sparse propagation environments. Ongoing work extends the approach to beamforming design and frequency-selective channels.

\section*{Acknowledgement}
The authors   gratefully acknowledge  Dr Rahif Kassab and Dr Dongzhu Liu for their contributions to the early stage of this project in terms of comprehensive literature review.

\bibliographystyle{ieeetr}	
\bibliography{Yusha_bib}

\end{document}